\documentclass[aps,prb,onecolumn,showpacs,showkeys]{revtex4}

\usepackage{epsfig}
\usepackage{amsmath}
\usepackage{amsfonts}
\usepackage{amssymb}
\usepackage{slashed}
\usepackage{graphicx}
\usepackage{subfigure}
\usepackage[latin1]{inputenc}
\newcommand{\be}{\begin{equation}}
\newcommand{\ee}{\end{equation}}

\begin{document}

\title{Wave equation of the scalar field and superfluids}
\author{A. Naddeo\footnote{e-mail: \textit{naddeo@sa.infn.it}}$^{(1)}$ and G. Scelza\footnote{e-mail: \textit{Giovanni.Scelza@unige.ch}}$^{(2)}$}
\affiliation{(1) CNISM, Unit\`{a} di Ricerca di Salerno and
Dipartimento di Fisica ``E. R. Caianiello'', Universit\`{a} degli
Studi di Salerno, Via Salvador Allende, $84081$ Baronissi
(SA), Italy \\
(2) Universit\'{e} de Gen\`{e}ve, Departement de Physique
Theorique, $24$, quai E. Ansermet, CH-$1211$ Gen\`{e}ve,
Switzerland}


\begin{abstract}
The new formal analogy between superfluid systems and cosmology,
which emerges by taking into account the back-reaction of the
vacuum to the quanta of sound waves \cite{noi}, enables us to put
forward some common features between these two different areas of
physics. We find the condition that allows us to justify a General
Relativity (GR) derivation of the hydrodynamical equation for the
superfluid in a four-dimensional space whose metric is the Unruh
one \cite{Unruh}. Furthermore we show how, in the particular case
taken into account, our hydrodynamical equation can be deduced
within a four-dimensional space from the wave equation of a
massless scalar field.
\end{abstract}
 \maketitle
\section{Introduction}
Recently there has been a growing interest in developing analogue
models aimed to probe aspects of the physics of curved space-time
and of quantum field theory on curved space-time. Condensed matter
analogues have been mainly proposed, because of their conceptual
simplicity and experimental feasibility in the laboratory. They
appear particularly useful in order to simulate kinematical
properties of curved space-time. Such a research area is appealing
also because sometimes insights gained within the GR context could
help to understand aspects of the analogue model.

Following this line of thought different condensed matter systems
have been introduced as analogue models. In such a context
acoustics in flowing fluids, light in moving dielectrics or
quasiparticles in moving superfluids, have been shown to reproduce
some aspects of GR and cosmology \cite{volov,liberati}. They can
be conceived as laboratory toy models in order to make
experimentally accessible some features of quantum field theory on
curved space-time. The analogy between the motion of sound waves
in a convergent fluid flow and massless spin-zero particles
exposed to a black hole was first outlined in the seminal paper by
Unruh \cite{Unruh}. Since then, the search for an emergent
space-time has been extended to various media, such as
electromagnetic waveguides \cite{Unruh1}, superfluid helium
\cite{volov} and Bose-Einstein condensates (BEC) \cite{liberati1}.
Emergent space-time and gravity effects in superfluids are of
particular interest. Indeed the extremely low temperatures
experimentally accessible allow in principle the detection of tiny
quantum effects, such as Hawking radiation, particle production
and quantum back-reaction \cite{volo,fischer1}. In particular,
liquid helium II offers the possibility of an experimental control
of the value of the speed of first sound, which corresponds to
thermal phonons. That may be achieved, for example, by fixing
temperature and varying pressure within a wide range (up to about
2.5 MPa) below the lambda transition point \cite{helium2}. On the
other hand, Bose-Einstein condensates made of cold atoms in
optical lattices are very promising because of the high degree of
experimental control \cite{optical}. Indeed such systems have been
proposed to mimic an expanding Freedman, Robertson, Walker (FRW)
universe \cite{wei}, where the behavior of quantum modes has been
reproduced by manipulating the speed of sound through external
fields via Feshbach resonance techniques \cite{Bar}. The key point
of such a finding is the relation $c_{s}^{2}\propto a_{s}$ between
the propagation speed $c_s$ and the $s$-wave scattering length
$a_s$ of the atoms of the condensate \cite{liberati1}. In this way
the value of $c_{s}$ can be changed at will by varying the
scattering length in a sufficiently slow manner by means of
suitable external fields. That happens without violating the basic
assumptions made in deriving the Gross-Pitaevskii equation
\cite{optical}, which describes BEC, and makes some of the
predictions of semiclassical quantum gravity and cosmology
testable in the laboratory \cite{Bar}.

In a recent paper \cite{noi} a new analogy between superfluids and
cosmology has been presented, which relies on the depletion of the
mass density $\rho$ in the superfluid due to thermal phonons. That
is the back-reaction of the vacuum to the quanta of sound waves.
Under these new conditions, the free energy is written as the sum
of two contributions: the energy of the quantum vacuum and the
free energy of the ``matter'', the phonons. Then, it is possible
to show \cite{volo} that
\begin{equation}\label{eqn:dyn}
  F(T,\rho)=F_{vac}+F_{mat}=F_{vac}-P_{mat}=\varepsilon(\rho)-\mu\rho-\frac{1}{3}\varepsilon_{mat}(\rho),
\end{equation}
where $\varepsilon-\mu\rho=\varepsilon_{vac}$ and
$\varepsilon_{mat}$ are, in this order, the \emph{energy density
of the quantum vacuum} and the \emph{energy density of the gas of
thermal phonons} (radiation energy). By expanding the free energy
$F$ in terms of $\delta \rho=\rho-\rho_0$ and $\delta
\mu=\mu-\mu_0$ (where $\rho_0$ and $\mu_0$ are the equilibrium
density and the chemical potential at $T=0$, $\rho=\rho(T\neq 0)$
and $\mu=\mu(T\neq 0)$) and then by minimizing over $\delta \rho$
the following hydrodynamical equation\cite{volo} is obtained
\begin{equation}\label{eqn:fir}
  \frac{\delta \rho}{\rho}=-\frac{\varepsilon_{mat}}{\rho c_s^2}\biggl(\frac{1}{3}+
\frac{\rho}{c_s}\frac{\partial c_s}{\partial \rho}\biggr).
\end{equation}
It is possible to show \cite{noi} that, under the hypothesis
$\frac{\delta\rho}{\varepsilon_{mat}}<<1$, Eq. (\ref{eqn:fir}) can
be cast in a form which looks very similar to the cosmological
fluid one (see Eq. (\ref{eqn:fluido}) for a comparison):
\begin{equation}\label{eqn:flu}
  \frac{d\rho}{dc_s}\simeq
-\frac{3}{c_s}\biggl(\rho-3
c_s^2\frac{\rho\,\delta\rho}{\varepsilon_{mat}}\biggr),
\end{equation}
$c_s$ being the phonon speed. Indeed in the cosmological context,
the Friedman fluid equation \cite{wei}
\begin{equation}\label{eqn:fluido}
  \frac{d\,\rho}{d\,a}=-\frac{3}{a}\biggl(\rho+\frac{p}{c^2}\biggr),
\end{equation}
where $a=a(t)$ is the scale parameter of the Universe, is obtained
starting from the Einstein equations by means of the condition
\begin{equation}\label{eqn:covdev}
  D_{\nu}T_0^{\phantom{\mu}\nu}=D_{\nu}T^{0\nu}=0,
\end{equation}
where $D_{\nu}$ is the covariant derivative.\\
Let us notice that the correspondence between the above equations
(\ref{eqn:flu}) and (\ref{eqn:fluido}) is
\begin{equation}\label{eqn:csan}
  c_s\leftrightsquigarrow a.
\end{equation}
We point out that the result (\ref{eqn:fir}), from which we deduce
(\ref{eqn:flu}), can be obtained through the analysis of a
classical hydrodynamic equation made by Stone \cite{stone} for
which the Unruh metric \cite{Unruh} holds. Furthermore the
conclusion (\ref{eqn:csan}) is a crucial one in that leads us to
derive in an alternative way the effective metric for the
superfluid.

Such a derivation is the aim of this letter. Indeed by exploiting
the mathematical analogy between the propagation of sound in a
nonhomogeneous potential flow and the propagation of a scalar
field in a curved space-time, in full analogy with Ref.
\cite{stone}, we will show that it is possible to introduce an
action $S$ and an energy-momentum tensor $T_{\mu \nu }=\frac{2}{%
\sqrt{-g}}\frac{\delta S}{\delta g^{\mu \nu }}$ in such a way that
the conservation law $D_{\nu }T^{\mu \nu }=0$ is satisfied,
$D_{\nu }$ being the usual GR covariant derivative.

In Section 2, following these steps, Eq. (\ref{eqn:flu}) will be
recovered and the formal analogy between GR and fluid dynamics
introduced in Ref.\cite{noi} will be fully exploited. Finally, in
Section 3 we summarize our conclusions and outline some
perspectives and open problems.
\section{The four-dimensional analogy}
In this Section we fully exploit the analogy between the
propagation of sound in a nonhomogeneous potential flow and that
of a scalar field in a curved space-time by carrying out an
analysis similar to the one developed by Stone \cite{stone} or
Fedichev and Fischer\cite{fed}. There it is shown how it is
possible to describe a perfect, irrotational fluid, within a
four-dimensional formalism, through the equation $\Box\Phi=0,$
where $\Phi$ is a scalar field, by means of the Unruh metric.

Let us start by finding an effective metric for the superfluid.
The key point of our reasoning is as follows. By looking at the Unruh metric \cite{Unruh}: $ds_{UN}^{2}=\frac{\rho \left( t,\overrightarrow{x}\right) }{c_{s}\left( t,%
\overrightarrow{x}\right) }\left\{ -\left[ \left( c_{s}\left( t,%
\overrightarrow{x}\right) \right) ^{2}-\left( \overrightarrow{v}\left( t,%
\overrightarrow{x}\right) \right) ^{2}\right] dt^{2}-2\overrightarrow{v}%
\left( t,\overrightarrow{x}\right) \cdot dtd\overrightarrow{x}+d%
\overrightarrow{x}^{2}\right\}$, where $\overrightarrow{v} \left(
t,\overrightarrow{x}\right)$ is the physical velocity of the
superfluid with respect to the laboratory, it is possible to
deduce a \textit{minkowskian acoustic metric} in the case
$\overrightarrow{v} \left( t,\overrightarrow{x}\right)=0,$ i.e.
the case of an inner observer. By means of a \emph{conformal
transformation}\cite{haw}, such a metric can be written as
\begin{equation}\label{eqn:met}
  [g_{\mu\nu}]=\frac{1}{\rho(t)}\begin{pmatrix}1&0&0&0\\
0&-c_s^{-2}(t)&0&0\\
0&0&-c_s^{-2}(t)&0\\
0&0&0&-c_s^{-2}(t) \end{pmatrix}.
\end{equation}
Let us notice that the density $\rho$ is a function of
the time alone in order to describe an homogeneous fluid.\\
Now, within the GR context the cosmological fluid equation
(\ref{eqn:fluido}) is derived starting from the Einstein equations
which do not hold for a quantum fluid. Then, in order to satisfy
the condition (\ref{eqn:covdev}) also for the analogue superfluid system
under study we need to proceed in a different way.\\
Our strategy is the following. Let us suppose that the
hydrodynamical equation (\ref{eqn:flu}) is derived by a wave
equation of the kind
\begin{equation}\label{eqn:box}
  \Box\Phi=\frac{1}{\sqrt{-g}}\partial_{\mu}(\sqrt{-g}g^{\mu\nu}\partial_{\nu}\Phi)=0,
\end{equation}
where $\Phi=\Phi(t,r,\theta,\phi)$ is some massless scalar field
and $g_{\mu\nu}$ is the metric tensor (\ref{eqn:met}), with
$g=Det(g_{\mu\nu})$.\\
We explicitly note that for a pure radiation field, that is the
case considered in (\ref{eqn:box}), it is
$\frac{p}{c^2}=\frac{\rho}{3}$, then Eq. (\ref{eqn:fluido}) can be
written as $\frac{d\rho}{da}=-\frac{4}{a}\rho$. Remembering our
hypothesis about $\frac{\delta\rho}{\varepsilon_{mat}}$, we can
rewrite (\ref{eqn:flu}) as $\frac{d\,\rho}{d\,c_s}\simeq
-\frac{3}{c_s}\rho$
 and, then, the formal similitude between Eqs. (\ref{eqn:flu}) and (\ref{eqn:fluido})
is preserved.\\
The result is:
\begin{equation}\label{eqn:wav}
  \Box \Phi=0\Rightarrow
  \frac{d\rho}{dc_s}=-\frac{\rho}{c_s}\Biggl(3-\frac{c_s}{\dot{c_s}}\frac{\partial_{t,t}\Phi}{\partial_t\Phi}+
\frac{c_s^3}{\dot{c_s}\partial_t\Phi}\biggl(\partial_{\varphi,\varphi}\Phi+
\partial_{\theta,\theta}\Phi+\partial_{r,r}\Phi\biggr)\Biggr)=0.
\end{equation}
Here, $\dot{c_s}$ is the time derivative of the sound velocity,
and $\partial_x$ and $\partial_{x,x}$ are the first and second
order partial derivatives with respect to the variable $x$. Now,
by making a simple comparison among Eqs. (\ref{eqn:wav}) and
(\ref{eqn:flu}), we can also find (with arbitrary parameters) the
suitable
expression for the field $\Phi(t,r,\theta,\phi).$\\
In order to analytically solve such an equation, we make the
following simplifying assumptions:
\begin{itemize}
  \item $\Phi$ is a function of the time alone: $\Phi=\Phi(t)$;
  \item the sound velocity is equal to $c_s(t)=\gamma t^{\frac{1}{2}}$,
  so that $c_s\dot{c}_s=\frac{\gamma^2}{2}$, $\gamma$ being a constant
\footnote{let us remember that
  the cosmological analogue $a(t)\propto t^\frac{1}{2}$ holds for an Universe \emph{radiation
  dominated}.};
  \item the quantity $\zeta=\frac{\delta\rho}{\varepsilon_{mat}}$
  is assumed to be a constant.
\end{itemize}
In this way, by means of a direct comparison with Equation
(\ref{eqn:flu}), Equation (\ref{eqn:wav}) can be easily solved and
a simple solution for the field $\Phi$ is given by
\begin{equation}\label{eqn:fi}
  \Phi(t)=\sqrt{2}\,\frac{e^{k\,t}}{k},
\end{equation}
where $k=\frac{9}{2}\gamma^2\zeta.$ Such a field can be identified
with the sound field, which is expected to correspond to a quantum
coherent state of phonons \cite{stone}.

The next step to carry out in order to fully exploit the
similarity with GR is to find an an action $S$ from which to
deduce an energy-momentum tensor that could allow us to achieve
Eq. (\ref{eqn:flu}). Proceeding in analogy with Reference
\cite{stone}, let us introduce an action $S$ defined as:
\begin{equation}\label{eqn:act}
  S=\int
  d^4x\frac{1}{2}\sqrt{-g}g^{\mu\nu}\partial_{\mu}\Phi\partial_{\nu}\Phi=\int
  d^4x\sqrt{-g}\mathcal{L},
\end{equation}
where the sound field $\Phi$ is defined through Eq.
(\ref{eqn:box}). Let us remember explicitly that such an action
gives rise to Eq. (\ref{eqn:box}) by setting equal to zero the
variation of the action with respect to $\Phi$, i.e.
$\delta_{\Phi} S=0$ (see for instance Ref. \cite{birda}). It is
well known \cite{wei,lan} that any action $S$ automatically
provides us with a covariantly conserved and symmetric
energy-momentum tensor
\begin{equation}\label{eqn:enimte}
  T_{\mu\nu}=\frac{2}{\sqrt{-g}}\frac{\delta_{g}S}{\delta
  g^{\mu\nu}},
\end{equation}
where now $\delta_{g}S$ is the variation of the action with
respect to the metric. In this way we find
$$T^{\mu\nu}=\partial^{\mu}\Phi\partial^{\nu}\Phi-\frac{1}{2}g^{\mu\nu}
(g^{\alpha\beta}\partial_{\alpha}\Phi\partial_{\beta}\Phi),$$
that, in the case under study, takes the form:
\begin{equation}\label{eqn:tenenimp}
  [T^{\mu\nu}]=e^{2k\,t}\begin{pmatrix}\rho(t)^2&0&0&0\\
0&c_s(t)^2\rho(t)^2&0&0\\
0&0&c_s(t)^2\rho(t)^2&0\\
0&0&0&c_s(t)^2\rho(t)^2
\end{pmatrix},
\end{equation}
where the result of Eq. (\ref{eqn:fi}) has been introduced.\\
The condition $\delta_g S=0,$ gives rise to the following crucial
one \cite{lan}:
\begin{equation}\label{eqn:cov}
  D_{\nu}T^{\mu\nu}=T^{\mu\nu}_{\phantom{\mu\nu};\nu}=\frac{\partial
  T^{\mu\nu}}{\partial x^{\nu}}+\Gamma^{\mu}_{\nu\sigma}T^{\sigma\nu}+
\Gamma^{\nu}_{\nu\sigma}T^{\mu\sigma}=0,
\end{equation}
where it appears clearly that $D_{\mu}$ is the usual GR definition
of the covariant derivative. In particular, from the equation
\begin{equation}\label{eqn:cov1}
  T^{0\nu}_{\phantom{\mu\nu};\nu}=0
\end{equation}
we re-obtain the hydrodynamical equation (\ref{eqn:flu}). That is
a statement of the correctness of our hypotheses and a proof of
the formal analogy between GR and superfluid dynamics introduced
in Ref. \cite{noi}.

Furthermore let us notice that, starting from:
\begin{equation}\label{eqn:tenenimp_2}
  [T^\mu_{\phantom{\mu}\nu}]=e^{2k\,t}\begin{pmatrix}\rho(t)&0&0&0\\
0&-\rho(t)&0&0\\
0&0&-\rho(t)&0\\
0&0&0&-\rho(t)
\end{pmatrix},
\end{equation}
the energy density and the pressure of the massless scalar field
$\Phi$ are derived and take the form (we follow the definition of
Kolb and Turner \cite{kolb}):
$$\rho_{\Phi}=T^0_{\phantom{\mu}0}=e^{2k\,t}\rho(t)$$
and
$$p_{\Phi}=\frac{1}{3}T^i_{\phantom{\mu}i}=-e^{2k\,t}\rho(t).$$
Then, finally, we automatically get the relation
\begin{equation}\label{eqn:infl}
  \rho_{\Phi}=-p_{\Phi},
\end{equation}
which coincides with the equation of state for the inflaton field
in an inflationary universe \cite{hobson}. We plan to further
investigate this topic and to clarify how inflationary dynamics
can be mimicked in the laboratory in a future publication
\cite{noinew}.
\section{Conclusions and perspectives}
In conclusion, by means of the field equation (\ref{eqn:wav}) we
have rewritten the hydrodynamical equation (\ref{eqn:flu}) within
a four-dimensional framework. That allowed us to make a direct
comparison among superfluid dynamics and GR. Following Stone
\cite{stone} we were able to introduce an action $S$ and an
energy-momentum tensor $T^{\mu\nu}$ for the sound field $\Phi$.
That allowed us to derive the relevant hydrodynamical equation
(\ref{eqn:flu}) from the condition $D_{\nu}T^{\mu\nu}=0$ even if
superfluid dynamics theory lacks general covariance. In this way
the analogy between GR and superfluid theory can be made very
transparent. We stress that within the new conditions considered
here, i. e. the back-reaction of the vacuum to the quanta of sound
waves, we are faced with a sound speed $c_s$ depending on the
time. In order to make Equation $\Box \Phi=0$ equivalent to
Equation (\ref{eqn:flu}), the relation $c_s\propto
t^{\frac{1}{2}}$ has been found.

The relevance of finding similarities among different fields of
research and of building up analogue models in order to test
theories which otherwise could not be proven. In the particular
case we study in this letter, the paradigm of cosmology as a
research area where it is not possible to check hypotheses is
bypassed since we have a toy model, a superfluid system built in
the laboratory, where our predictions could be tested. Then, all
the formal similarities we find could give us new insights to
understand the nature of the universe. For instance, in the
context of superfluid systems the presence of thermal phonons
plays the same role as the matter in the universe. At this stage a
question seems to emerge as a speculation: within the cosmological
area, can we conceive the matter as a thermal
perturbation of the vacuum as for the superfluids?\\
\section{Acknowledgements}
We wish to thank Dr. F. Maimone for valuable discussions and
suggestions.

\end{document}